# Light-field-driven non-Ohmic current and Keldysh crossover in a Weyl semimetal

R. Ikeda[1] *, H. Watanabe[2,1] †, J. H. Moon[3], M. H. Jung[3,1], K. Takasan[4], S. Kimura[2,1,5] ‡

[1]Department of Physics, Graduate School of Science, Osaka University, Toyonaka, Japan
[2]Graduate School of Frontier Biosciences, Osaka University, Suita, Japan
[3]Department of Physics, Sogang University, Seoul, Republic of Korea
[4]Department of Physics, The University of Tokyo, Tokyo, Japan
[5]Institute for Molecular Science, Okazaki, Aichi, Japan

**In recent years, coherent electrons driven by light fields have attracted significant interest in exploring novel material phases and functionalities[1–4]. However, observing coherent light-field-driven electron dynamics[5,6] in solids is challenging because the electrons are scattered within several ten femtoseconds in ordinary materials, and the coherence between light and electrons is disturbed. However, when we use Weyl semimetals[7–11], the electron scattering becomes relatively long (several hundred femtoseconds - several picoseconds), owing to the suppression of the back-scattering process[12]. This study presents the light-field-driven dynamics by the THz pulse (~1 ps) to Weyl semimetal $Co_3Sn_2S_2$[13–20], where the intense THz pulse of a monocycle electric field nonlinearly generates direct current (DC) via coherent acceleration without scattering[5,6] and non-adiabatic excitation[1,5,21,22] (Landau–Zener Transition). In other words, the non-Ohmic current appears in the Weyl semimetal with a combination of the long relaxation time and an intense THz pulse. This nonlinear DC generation also demonstrates a Keldysh crossover[23] from a photon picture to a light-field picture by increasing the electric field strength.**

Light exhibits a dual nature as a photon and a light field. In light-matter interactions, the border of the photon and the light field is known as the Keldysh parameter[1,5,23,24] $\gamma(= \omega\sqrt{m\Delta}/|e|E)$, where $\omega$ is the frequency of light, $m$ is the mass of the electron, $\Delta$ is the band gap energy, $e$ is the electron charge, and $E$ is the electric field strength of the light. For $\gamma > 1$, light acts on matter as a photon, which can be attributed to a traditional particle (photon) picture. For $\gamma < 1$, light cannot be described as a photon as it behaves as an electric field, namely a light field. Typical examples are high-order harmonics generation (HHG)[25–30], the Landau–Zener transition[1,5,21,22], and the Floquet state[31-34]. Non-equilibrium phenomena induced by a light field have attracted significant attention for exploring novel material functionalities and quasi-equilibrium phases. Hence, the investigation of light-field-driven phenomena is currently necessary. In addition to the Keldysh parameter lower than 1 ($\gamma < 1$), coherence between the light field and electrons is essential for observing light-field-driven phenomena. Coherence is easily broken by electron-electron and/or electron-phonon scattering. As the electron scattering time in ordinary materials is very short (in the order of 10 fs), ultrashort pulses of less than 10 fs have been used to coherently drive electrons in materials[5,6]. To obtain ultrashort pulses of 10 fs or less, light with energies in the mid-infrared range (~100 THz) or higher is required. However, high-energy light can cause real electron excitations and prevent coherent driving. To realize coherent driving of electrons in materials, it is ideal to drive electrons in materials with a long electron scattering time using a low-energy light field[35], i.e., a THz electric field. According to the Keldysh parameter $\gamma$, ultra-intense light (~$10^9$ V/m) is required to achieve a light field picture at high frequencies, such as near-infrared[5]; however, using longer wavelengths, such as the THz wave (~ 1 ps), the light field picture ($\gamma < 1$) can be easily achieved even in a weak electric field (~$10^5$ V/m). Therefore, we performed a THz-field drive in a Weyl semimetal, with a long electron scattering due to spin-momentum rocking, maintaining a long coherence between light and electrons. Therefore, it is expected that a light-a, THz pulse $E_S$ reflected/emitted from the sample. b, Fourie transform (FT) spectrum of the THz pulse reflected/emitted from the sample. c, Dependence of time integration for THz pulses on electric field strength. Time integrations are performed over a range of -2–4 ps. d, Electric field waveform for the AC component $E_{AC}$ of the THz pulse reflected/emitted from the sample as a function of the introduced electric field strength. e, FT spectra of AC components. The FT spectra of Ref (Cu plate) is detected by a gallium phosphorate (GaP) crystal. f, Electric field waveform for the DC component $E_{DC}$ of the THz pulse reflected/emitted from the sample. g, FT spectra of the DC components. h, Schematic of DC generation by a monocycle intense THz pulse drive.

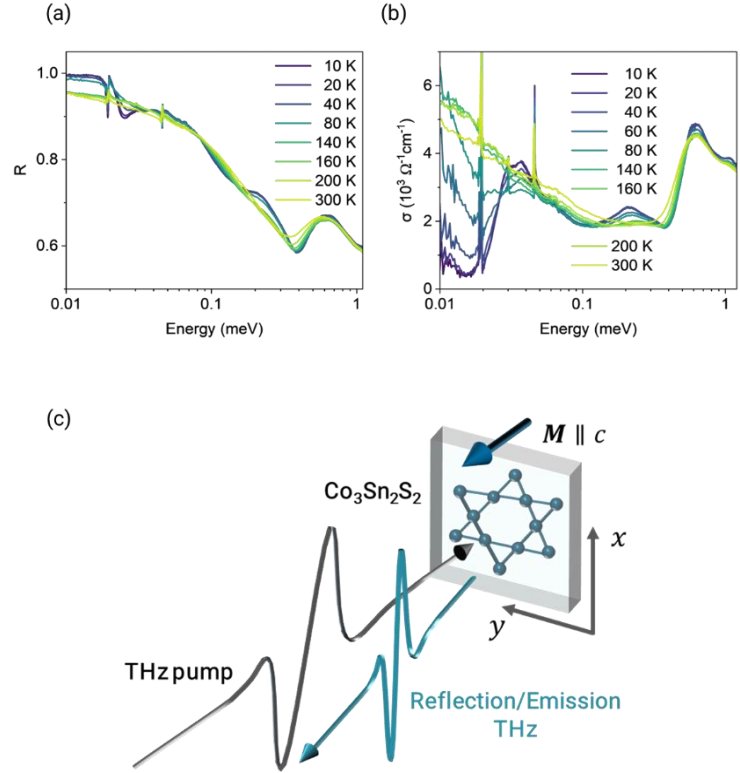

**Fig.1| Reflectivity (R) and optical conductivity (σ) spectra of $Co_3Sn_2S_2$ and experimental configuration.**
a, R spectra in the THz and IR as a function of temperature. b, Temperature-dependent σ spectra. At low temperatures, a pseudogap appears at photon-energies below 30 meV, and the Drude tail of the carriers is visible at photon energies below 15 meV. c, Experimental configuration. The intense THz pulse (black pulse) is irradiated onto the c-axis plane along the magnetization direction $M$. THz pulse reflected/emitted (green pulse) by sample ($Co_3Sn_2S_2$) is detected by THz-TDS.

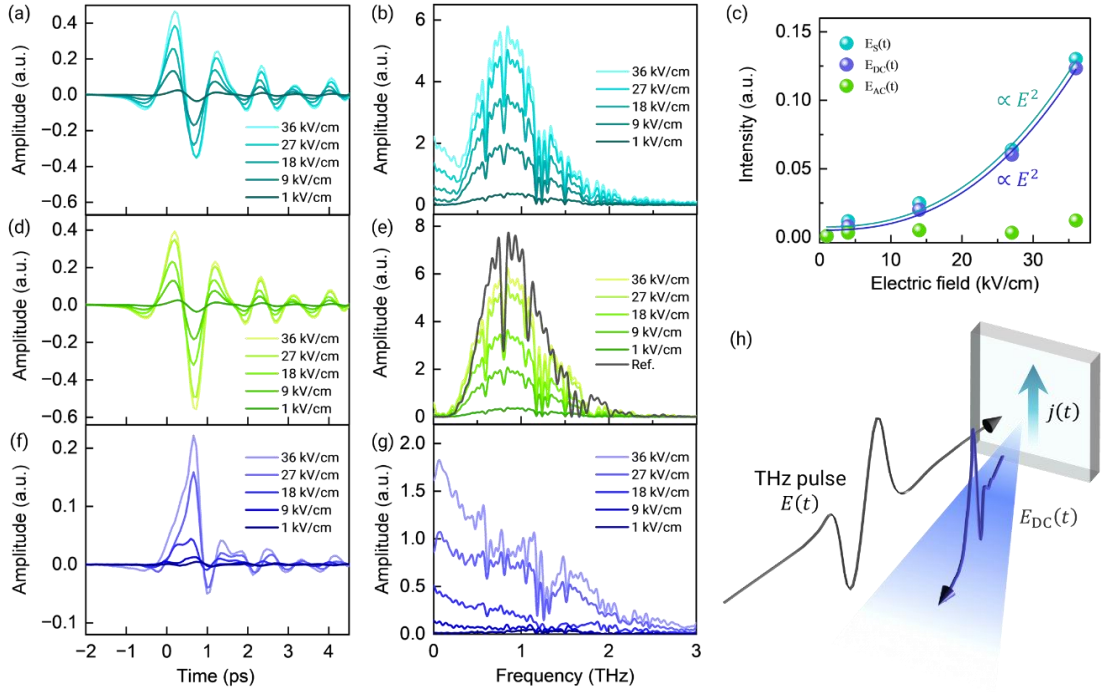

**Fig.2| Electric field strength of THz pump/reflective THz-TDS in $Co_3Sn_2S_2$.**

field drive can be realized using a THz waves. Furthermore, Weyl semimetals, which have a Dirac point only tens meV away from the Fermi energy[12], have a long coherence time and inhibit real excitation; therefore, they are ideal for light-field drives. Therefore, the THz wave is the ideal light source for observing of Weyl fermion dynamics driven by light fields.

This study shows the light-field-driven electrodynamics of the Weyl semimetal $Co_3Sn_2S_2$ (CSS)[13–20] in the THz frequency (~ 1ps). CSS has a relatively long electron scattering time of 0.1–10 ps, a considerably slower timescale than the previously reported graphene[5] and organic material[6]. As the time scale is comparable to a monocycle THz pulse, carrier dynamics by coherent acceleration can be observed. Consequently, we demonstrate THz-pulse-driven direct current (DC) generation in a CSS via coherent acceleration and non-adiabatic excitation (Landau–

Zener transition[1,5,21,22]), which is completely different from Ohm's law. The nonlinearity of the DC with respect to the THz-electric-field strength indicates that the Landau–Zener transition, which is specific to a light field, becomes more pronounced as the electric field strength increases. This indicates a Keldysh crossover[23] with increasing electric field strength of the THz pulse.

First, the carrier dynamics and electronic structure of CSS are described. Figures 1(a) and 1(b) show the temperature dependence of the reflectivity (R) and optical conductivity (σ) spectrum, derived from the Kramers–Kronig analysis of the R spectrum, in the c-plane of CSS, respectively. The R spectrum exhibits a strong temperature dependence representing electronic structure modulation with temperature due to the phase transition from a normal semimetal to a Weyl semimetal with decreasing temperature across $T_C$=177 K[13–20]. In the Weyl semi-metallic phase, a pseudogap appears at a photon energy of approximately 30 meV, and a renormalized Drude component emerges below 15 meV. The Drude component has a relaxation time as long as 1 ps, which satisfies the condition for coherent acceleration by a THz pulse. Subsequently, the coherently acceleration was measured using a THz pulse, as shown in Fig. 1(c). THz pulse was irradiated onto a single crystalline CSS in the c-plane, and the reflection/emission spectra from CSS were measured using THz-time-domain-spectroscopy. The THz pulse has a linear polarization (x polarization) with a maximum field strength of 36 kV/cm and a center frequency of 0.8 THz. In this study, we performed the experiments at a temperature of approximately 7 K, which is significantly lower than the Curie temperature.

Figures 2(a) and 2(b) show the electric field waveform $E_S$ of a THz pulse reflected/emitted by the CSS and its Fourier transform (FT) spectrum, respectively, depending on the incident electric field strength. Figure 2(b) shows a peak structure centered at 0.8 THz, similar to the reference (Cu plate) spectrum, and, a Drude-like spectral structure below 0.2 THz only at high electric field incidence, indicating that high light fields generate a DC. In the general case of THz-TDS, as the THz pulse has a symmetric amplitude, the reflected THz pulse $E_S$ is symmetric, and the time integral of its waveform is zero. However, in this case, the time integral of $E_S$ at high light fields is not zero as shown in Fig. 2(c), and, is proportional to the square of the electric field strength ($\propto E^2$). $E_S$ can be decomposed into two components, $E_{DC}$ for the DC component and $E_{AC}$ for the remaining current (AC) component, in which the shape and intensity of the $E_{AC}$ component is evaluated as a waveform obtained at the lowest laser intensity and the multiplication of the incident laser intensity relative to the lowest one, respectively. The electric field waveforms of $E_{DC}$ and $E_{AC}$ are shown in Figs. 2(d) and 2(f), respectively, and their FT spectra are shown in Figs. 2(e) and 2(g), respectively. The FT spectra of $E_{AC}$ shown in Fig. 2(e) are similar to the reflection spectrum from a reference material (Cu plate) originating from a simple reflection, indicating high reflectivity due to the metallic character shown in Fig. 1(a).

On the other hand, the DC component has a positively biased asymmetric-electric-field waveform, as shown in Fig. 2(f). The FT spectrum of the DC component shown in Fig. 2(g) appears to be a broadband Drude-like spectrum below 3 THz. The integrated intensity is proportional to the square of the incident electric field amplitude ($E_{DC} \propto E^2$), which is consistent with the behavior of $E_S$. This result suggests that the asymmetric emitted electric waveform originates from the DC component, i.e., the observed asymmetric emission and DC generation are driven by a monocycle-intense THz pulse (depicted in Fig. 2(h)).

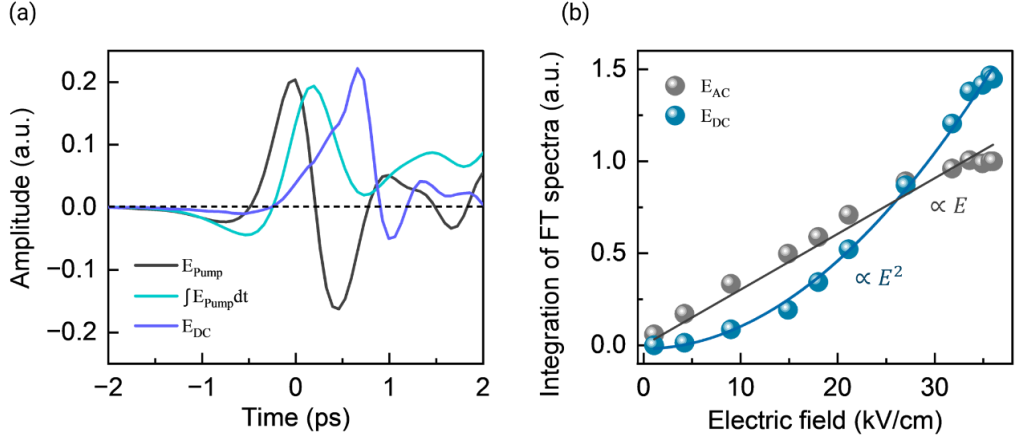

**Fig.3| THz coherent acceleration-induced DC and incident electric field dependence.**
a, Comparison of the electric field waveforms. The DC component $E_{DC}$, intense THz pulse, and time integration of the intense THz pulse, are indicated by the purple, black, and green curves, respectively. b, Electric field strength dependence of the integrated FT spectra for $E_{DC}$ and $E_{AC}$. $E_{AC}$ is proportional to the electric field strength, as indicated by the gray fitting line, and $E_{DC}$ is proportional to the square of the electric field strength, as indicated by the blue fitting curve.

Next, the origin of DC generation is discussed. Assuming that the Ohm's law ($j(t) = \sigma E(t)$, where $j(t)$ is the current and $\sigma$ the conductivity) is dominant, when a symmetric THz pulse waveform isotropically accelerates the electrons in a material, no DC components appear. However, the DC components are visible as a THz pulse, as shown in Fig. 2(g). If no scattering occurs within the THz pulse acceleration time, the DC component $j(t)$ can be generated by the incident THz pulse $E(t)$, as expressed by Eq. (1).

$$j(t) = \frac{ne^2}{m} \int E(t)\, dt, \qquad (1)$$

where $n$, $e$, and $m$ denote the carrier density, electron charge, and electron mass, respectively. The differential current generates THz emissions as follows:

$$E_{DC}(t) \propto \frac{d\boldsymbol{j}}{dt}. \qquad (2)$$

Therefore, if carrier density is preserved, the DC electric field of the THz emission is proportional a, Schematic of the Landau–Zener transition in the Weyl band. b, Schematic crossover diagram from a photon picture ($\gamma > 1$) to a light-field picture ($\gamma < 1$) at the Keldysh line. The crossover is caused by the increase in the electric field strength (blue arrow).

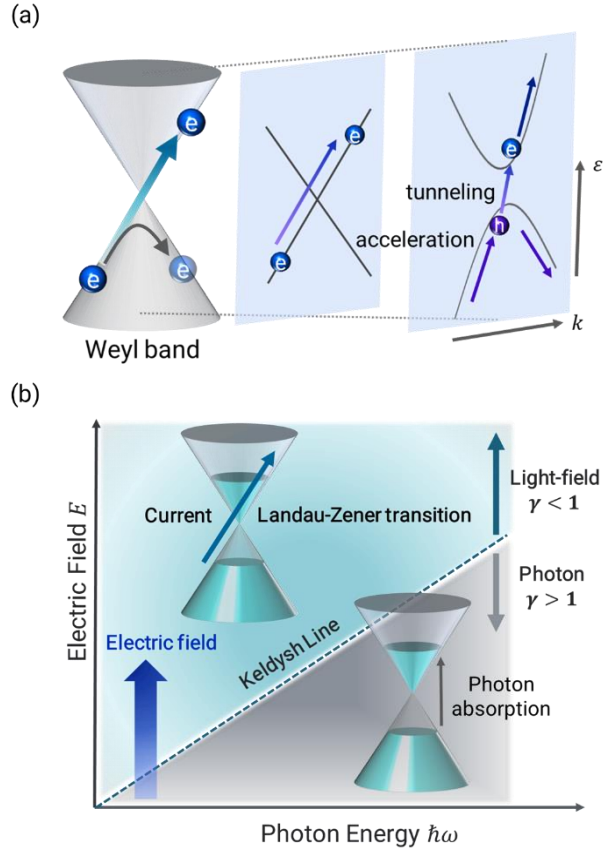

**Fig.4| Landau–Zener transition in Weyl band and Keldysh line crossover.**

to the incident THz pulse ($E_{DC}(t) \propto E(t)$); however, this is not consistent with the experimental results, i.e., $E_{DC}(t)$ does not follow $E(t)$ but follows $\int E(t)\,dt$, as shown in Fig. 3(a), indicating a coherent acceleration[6]. To understand this inconsistency, the carrier density is changed with time, i.e., Eq. (1) can be rewritten as follows:

$$\frac{dj}{dt} = \frac{e^2}{m}\left(\frac{\partial n}{\partial t}\right)\int E(t)\,dt + \frac{e^2}{m}nE(t). \qquad (3)$$

As shown in Fig. 3(a), the second term may only make a minor contribution; however, the first term is dominant because $E_{DC}(t)$ almost follows $\int E(t)\,dt$, i.e., both $\int \boldsymbol{E}(t)dt$ and $E_{DC}$ have asymmetric positively biased waveforms with well-matched amplitude signs. This suggests that carriers are generated by the high electric field. To investigate the origin of the carrier excitation, we focus on the electric-field-strength dependence of $E_{DC}(t)$, as shown in Fig. 3(b). $E_{AC}$ exhibits linear dependence with the incident electric field amplitude owing to a simple metallic reflection at the sample. In contrast, the area of the FT spectra of $E_{DC}$ is proportional to the square of the incident electric field. One possible origin of such nonlinear phenomena is two-photon absorption; however, the energy of the Weyl points from the Fermi level of CSS (several tens of millielectronvolts) cannot be excited by a two-photon process of approximately 4 meV. Multiphoton excitation is possible; however, multiphoton absorption is unlikely because of the small transition probability. In addition, the experimental $E^2$-law does not fit the multiphoton process.

Following the coherent acceleration by the intense THz pulse, another possible mechanism for carrier excitation is the Landau–Zener transition[1,5,21,22] caused by the light-field drive. As shown in Fig. 4(a), the Landau–Zener transition is caused by quantum tunneling, in which the electron is accelerated by a light field, resulting in a large momentum change. The current associated with the Landau–Zener transitions in band structures with linear dispersion, such as graphene and Weyl semimetals, is reportedly proportional to the $d-1$ power ($d$: dimension) of the electric field amplitude[36,37]. Owing to the three-dimensional Weyl semimetal CSS, the current intensity associated with the Landau–Zener transition in the material is expected to be proportional to the square of the electric field strength ($\propto E^2$)[37]. As shown in Fig. 3(b), the electric field strength dependence of $E_{DC}$ obeys the $E^2$-law, which agrees well with theoretical predictions. The DC is the origin of the coherent acceleration, and the DC intensity is proportional to the square of the THz electric field intensity. Therefore, the carrier excitation is attributed to the Landau–Zener transition. The Landau–Zener transition is a phenomenon specific to light-field pictures, therefore, as shown in Fig. 4(b), the crossover from a photon picture at a lower electric field to a light-field picture at a higher electric field can be observed by increasing the electric field strength of the incident light. The Keldysh parameter can be described by the following equation[5].

$$\gamma = \frac{\Delta}{2\hbar\Omega_R}. \quad (4)$$

$$\Omega_R = \frac{v_F |e| E}{\hbar\omega}, \quad (5)$$

where $\Omega_R$ is the Rabi frequency, $\omega$ is the frequency of light, $\Delta$ is the band gap energy, $v_F$ is the Fermi velocity of the electron, $e$ is the electron charge, and $E$ is the electric field strength of light. The band gap energy $\Delta$ corresponds to twice the Fermi energy due to the gapless structure of the Weyl band. Assuming $\Delta=30$ meV, $v_F = 1.0 \times 10^6$ m/s and $\hbar\omega = 4$ meV, the Keldysh parameter is 1 at $E = 6$ kV/cm. As shown in Fig.4(b), a light-field (photon) picture is dominant for $\gamma < 1$, and a photon picture is dominant for $\gamma > 1$. The electric field strength changes from 1 to 36 kV/cm, for which the dependence is measured in this experiment, corresponding to a change in the Keldysh parameter $\gamma$ from 4 to 0.1. At electric fields higher than $E = 6$ kV/cm ($\gamma \sim 1$), the DC component $E_{DC}$ increases significantly, as shown in Fig. 2(c). The appearance of the light-field picture observed in this study suggests a Keldysh crossover[23] in a Weyl semimetal.

As shown in Fig. 2(g), the DC generation also suggests that the THz pulse breaks the spatial-inversion symmetry of the electron system in a non-equilibrium state[38–40]. As a change in the system symmetry leads to a change in topology, the possibility of ultrafast topology control[31,33,41,42] by an intense THz pulse may be feasible.

In this study, we observed DC generation from a Weyl semimetal $Co_3Sn_2S_2$ by irradiating it with an intense monocycle THz pulse. The DC amplitude proportionally increases to the square of the incident electric field amplitude, suggesting that the intense THz pulse generates a non-Ohmic current in the Weyl semimetal and that its origin is the coherent acceleration by the light field drive and electronic excitation by the Landau–Zener transition. The nonlinear DC generation driven by a monocycle intense THz pulse suggests the realization of a light-field picture across a Keldysh crossover from a photon picture. By combining a THz pulse with a low frequency and Weyl semimetals with a gapless structure, we observed the phenomenon of a light-field picture ($\gamma < 1$) even at an easily achievable light-field strength of several tens of kilovolts per centimeters. The long electron scattering time in Weyl semimetals allows the occurrence of the light-field-driven phenomena even on a picosecond time scale. These results show that the THz-pulse drive of Weyl fermions is a good platform for investigating the initial process of light-field-driven phenomena. Furthermore, the generation of DC by an intense THz pulse indicates a light-induced system symmetry change and is expected to enable the establishment of a new approach for ultrafast topological phase transitions. The proposed approach is expected to significantly contribute to the development and exploitation of novel non-equilibrium phases and functionalities of materials.

**Methods**

**Sample preparation**

Single crystals of $Co_3Sn_2S_2$ were grown using the melting method with a stoichiometric ratio of Co:Sn:S = 3:2:2. The stoichiometric mixtures sealed in an evacuated quartz tube were heated to 930 °C for 1 day and kept for 2 days. Then, they were slowly cooled down to 700°C over 7 days and kept for 2 days. The obtained crystals were well cleaved with a mirror-like surface perpendicular to the c-axis. The crystal structure was confirmed via X-ray diffraction to be a Kagome lattice of Co ions with a rhombohedral structure (space group R–3m)[13,14]. The Weyl semi-metallic electronic structure emerges below the Curie temperature Tc of approximately 177 K, owing to the time-reversal symmetry braking associated with the ordering of the internal magnetization[15-20].

**Wide-range reflectivity measurement**

Near-normal-incident reflectivity spectra were acquired over a wide photon-energy region of 10 meV–30 eV to ensure accurate Kramers–Kronig analysis (KKA). A Michelson-type rapid-scan Fourier spectrometer (JASCO Co. Ltd., FT/IR-6100) was used for the photon energy regions of 10 meV–1.5 eV with a specially designed feedback positioning system to maintain the overall uncertainty level less than ±0.5 % in the temperature range of 10–300 K[43]. The sample was evaporated in situ with gold to obtain absolute reflectivity. In the energy region of 2–30 eV, the reflectivity spectrum was measured only at 300 K using synchrotron radiation at beamline 3B of the UVSOR Synchrotron Facility[44]. To obtain optical conductivity spectra via the KKA of the reflectivity spectra, the spectra were extrapolated below 10 meV with the Hagen–Rubens function, and above 30 eV with a free-electron approximation $R(\omega) \propto \omega^{-4}$ [45].

**Intense THz pump/reflective(emission) THz-TDS**

We constructed an intense THz pump/reflective(emission) THz-TDS system to observe the response of bulk crystals when pumped with an intense THz pulse (Supplementary Information No. 1). A Ti:Sapphire laser (1.5 eV, 50 fs, 1 mJ, and 1 kHz) was used as the light source for the generation of the intense THz and the signal sampling pulses. The intense THz pulse irradiated onto the sample was generated from a lithium niobate crystal using a tilted pulse front scheme. The THz pulse reflected/emitted from the sample was measured using THz-TDS with a ZnTe crystal. The electric field strength of the THz pulse irradiated onto the sample was varied using wire-grid pairs. The sample was cooled using a He cryostat (MictostatHe, Oxford Instruments). A THz window, Tsurupica (Broadbald Inc.), was used for the cryostat. The experiment was performed with the sample cooled to 7 K, and a magnetic field was applied using a Ne-Fe-B permanent magnet.

*These authors contributed equally; r.ikeda@edu.k.u-tokyo.ac.jp

† These authors contributed equally; hwata@fbs.osaka-u.ac.jp

‡ These authors contributed equally; kimura.shin-ichi.fbs@osaka-u.ac.jp

**Acknowledgements**

This work was supported by JSPS KAKENHI of Japan (Grant No. 20H04453) and the National Research Foundation of Korea (NRF) (Grant Nos. 2020R1A2C3008044 and 2022R1A4A1033562). The work of K. T. was supported by JSPS KAKENHI Grant No.JP22K20350 and JST PRESTO Grant No. JPMJPR2256.

**Author contributions**

R.I. conceived the study. The THz-pump/reflective THz-TDS set-up was developed by R.I. and H.W. THz-pump/reflective THz-TDS measurements were performed and analyzed by R.I and H.W., and wide-range optical reflectivity spectra were measured and analyzed by S.K. and M.H.J. The sample was grown and characterized by J.M. and M.H.J. Further, R.I., H.W., K.T., and S.K. discussed the results. S.K. supervised the study. The manuscript was written by R.I., H.W., and S.K. with input from all authors.

**Corresponding authors**

Correspondence to R. Ikeda, H. Watanabe or S. Kimura.

**Competing interests**

The authors declare no competing interests.

# Supplementary Information for Light-field-driven non-Ohmic current and Keldysh crossover in a Weyl semimetal

R. Ikeda[1], H. Watanabe[2,1], J. H. Moon[3], M. H. Jung[3,1], K. Takasan[4], S. Kimura[2,1,5]

[1]Department of Physics, Graduate School of Science, Osaka University, Toyonaka, Japan
[2]Graduate School of Frontier Biosciences, Osaka University, Suita, Japan
[3]Department of Physics, Sogang University, Seoul, Republic of Korea
[4]Department of Physics, The University of Tokyo, Tokyo, Japan
[5]Institute for Molecular Science, Okazaki, Aichi, Japan

## S1. The intense THz pump/reflective THz-TDS optical system

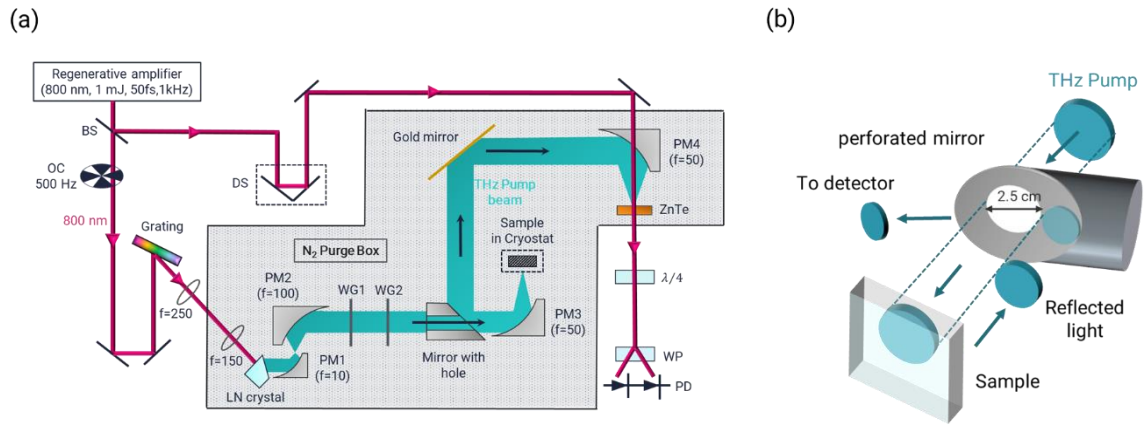

Fig.S1 The constructed intense THz pump/reflective THz-TDS optical system. (a) THz waves generated from the LN crystal are focused onto the sample, and the reflected light is reflected using a perforated mirror and focused onto the ZnTe crystal. (b) Details of a performed mirror.

  A schematic of the constructed intense THz pump/reflective THz-TDS optical system is shown in Fig. S1. The THz pulse was generated using the Cherenkov optical rectification technique using a LiNbO$_3$ nonlinear optical crystal, pumped by a Ti:sapphire regenerative amplifier (1 mJ, 1.55 eV, repetition rate of 1 kHz , pulse width of 50 fs). The generated THz beam was expanded by a factor of 10 using parabolic mirrors with focal lengths of 10 and 100 mm and then focused onto the sample using a parabolic mirror with a focal length of 50 mm. The spot size of the THz wave on the sample was approximately 500 μm. The THz waves reflected from the sample surface were reflected using a perforated mirror and guided to the detection system. The details are shown in Fig. S1(b). A hole with a 25 mm was drilled into the perforated mirror, and the pump THz beam was focused onto the sample after passing through this hole. By slightly tilting the sample, the path of the reflected light was shifted such that the THz beam was reflected at the mirror plane near the hole as shown in Fig. S1(b). A part of the Ti:sapphire regenerative amplifier (<1 µJ) was used for generating the sampling pulse for EO detection, and ZnTe crystals were used as the EO crystals. This experimental setup facilitates the observation of the ultrafast time evolution of the reflection spectrum instead of the transmission. Therefore, the ultrafast THz electric field response can be measured in bulk single crystals but not in thin-film samples.